# Secondary Ion Beams

*Klaus Knie*
GSI, Darmstadt, Germany

**Abstract**
Secondary ion beams are beams of particles produced by bombarding a production target with a primary beam of a stable nuclide (in most cases protons) or by fragmentation of heavy primary particles. These methods are used for short lived, instable particles which cannot be introduced into the ion source (or used as target). The secondary beam particles can be produced i) by bombarding a heavy, stable element in a second ion source (ISOL method), ii) by fragmentation of heavy beam particles, or iii) by creation of particle-antiparticle pairs (e.g. pions or antiprotons). Beams of decay products from secondary beams (e.g. muons or neutrinos from pion decay) can be considered as tertiary beams.

## 1    Introduction

A "standard" setup for a nuclear physics accelerator experiment consists of an ion source, an accelerator, a target and a detector system. A macroscopic amount of the element of choice is introduced in the ion source, ionised, the isobar of choice is selected and accelerated.

In case of protons, hydrogen comes from a commercial gas bottle. In the source the neutral atoms are ionized to H+ (e.g. linac2 at CERN) or H- (e.g. linac4) and accelerated by a succeeding accelerator. In case of heavy elements like uranium milligram amounts of metal powder are introduced into the source and ionized and $238U^{4+}$ is selected for further acceleration. The latter numbers are typical values used at GSI, Darmstadt.

After acceleration, a target is bombarded with the beam particles. There is a large variety of targets, but practically all consist of macroscopic amounts of material. Usually a target for a nuclear physics experiment is a foil of around 1 cm2 with a thickness of few mg/cm2. In case of colliders like the LHC, the beam itself can be considered as target, but again, macroscopic amounts of material are used for its production. The (non-relativistic) weight of the beam in the LHC is in the order of only one nanogram.

Mainly triggered by nuclear astrophysics, there is a huge interest of experiments with non-stable isotopes in order to get a better understanding for stellar nucleosynthesis. Especially in the r-process nuclides far from the valley of stability are involved. Therefore, either a radioactive beam, ideally with a high intensity, or a radioactive target, ideally with a high concentration of the respective radioisotope, is needed.

For very long-lived radioisotopes, this is mainly a matter of radiation protection (and budget, since large parts of the experimental setup need to be disposed afterwards). Just to give a few examples, macroscopic 14C (T1/2 = 5730 yr) or 59Ni (T1/2 = 108 kyr) have been produced already decades ago. For the production of superheavy elements 244Pu (T1/2 = 80 Myr) targets have been used. However, with decreasing half live of the isotope of interest not only radiation protection becomes more challenging, but the time available for measurements becomes shorter and shorter until no reasonable experiment is possible.

Besides of the interest on experiments with unstable nuclei there is a demand on experiments with unstable particles like pions, kaons, antiprotons or muons or with neutrinos. Here the problems are even more obvious: neither a few mg of pions can be introduced into an ion source in order to produce a pion beam for a few hours nor a target can be made out of antiprotons.

In order to overcome the problems mentioned above, secondary ion beams are a very powerful tool. The principles are depicted in fig. 1 and will discussed in detail in the next chapters. Fig. 1a) shows the standard setup for stable elements consisting of ion source, accelerator, ion optical system and experiment with target and detectors. In fig. 1b) the principle of the ISOL method (isotope separator online) is shown: Material in a second ion source (ion source is meant in a wider sense here) is continuously bombarded by a macroscopic primary beam and radioactive isotopes are produced there. Those are continuously extracted and accelerated by a second accelerator. An alternative to the ISOL method, the in-flight method, is shown in fig. 1c): A high energetic heavy ion beam, e.g. uranium, is fragmentated in a target. From the large number of different fragment nuclei those of interest are separated by a dedicated (and in some cases very sophisticated) ion optical system and guided to the experiments. The primary beam is dumped. Beside of fragment nuclei of the primary beam, one can also separate particles from particle-antiparticle pairs created in target, e.g. pions, kaons or antiprotons. To get a muon or neutrino beam, a secondary pion beam is produced which decays in muons and neutrinos, which can be considered a tertiary beam.

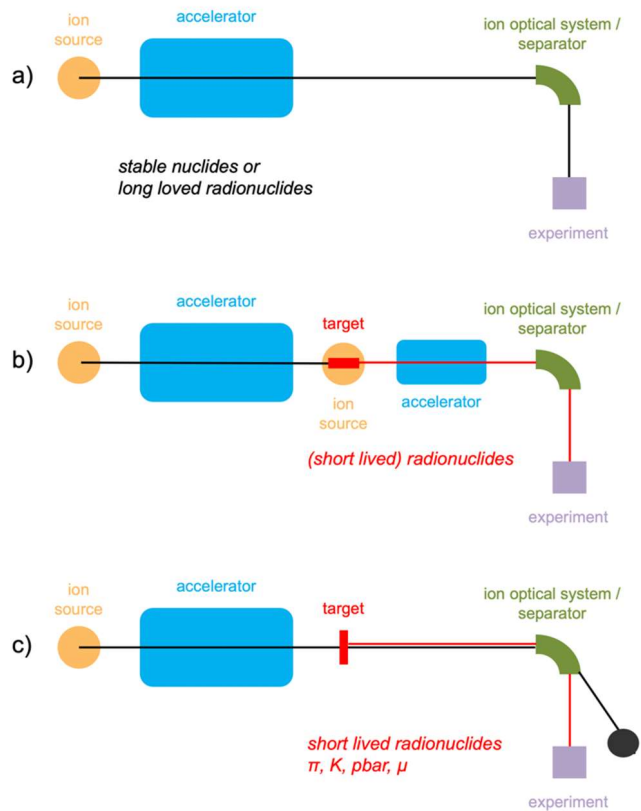

**Fig 1:** Setup for a standard nuclear physics experiment (a), the ISOL method (b), and the In-Flight method (c).

An additional challenge for both methods is the respective production target, because it has to withstand the very high primary beam intensity required for reasonable secondary beam intensities.

## 2    The ISOL method

The ISOL (Isotope Separator Online) is based on first experiments performed in Copenhagen already in the 1950s. In the 1960s CERN commissioned ISOLDE (Isotope Separator Online Device) [1], which is still in operation after a huge number of upgrades. Today, there is a number of ISOL facilities in operation, e.g. at ISAC at TRIUMF (Vancouver, Canada) [2], further facilities are planned or under construction.

The ISOL method can be considered as a stepwise process:

- A target made from a heavy element is bombarded by a primary beam with an energy in the order of one GeV and the radioisotope of choice (RI) is produced by nuclear reaction. Usually a proton beam is used, however, the ARIEL project presently under construction at TRIUMF will use electrons as primary beam particles [2]. With a production cross section $\sigma_{prod}$, a surface density of target nuclei $n_{target}$ and a number of primary particles $N_{prim}$, the production rate $dN_{RI}/dt$ of nuclide RI can be calculated to:

$$dN_{RI}/dt = \sigma_{prod} \cdot n_{target} \cdot dN_{prim}/dt \ .$$

  $\sigma_{prod}$ does not significantly change above primary beam energies in the order of 1 GeV. Therefore, the primary beam with 1.4 GeV for ISOLDE comes from the Proton Synchrotron Booster. Beam from the PS or even from the SPS is not necessary. At TRIUMF, the primaries have an energy of 500 MeV.

- The next step is a very challenging one. Although the radioisotope RI is produced now, one cannot use it, because it sticks in the target. It has to effuse through to the target surface and finally diffuse to the source's nearby ionization region. Whereas the first step has been pure nuclear physics, we are now in the realm of atomic physics and chemistry. The efficiency $\varepsilon_{e+d}$ depends strongly on the element and can reach from the percent to the ppm range. These processes can last in the order of seconds and dominate the time needed for the ISOL method by far. Therefore, one has to consider the decay of RI during this time by an efficiency $\varepsilon_{dec}$, which of course approaches 0 for half-lives which are very short compared to the diffusion time.

- Having RI finally in the ion source, it gets ionized with an efficiency $\varepsilon_{ion}$ and extracted.

- Besides of the wanted radioisotope RI a variety of other nuclides have been produced in the target and many of them made it into the source as well. Therefore, a separation of RI is necessary. A magnetic mass separation after the source is trivial. However, if the products $\sigma_{prod} \cdot \varepsilon_{e+d} \cdot \varepsilon_{dec} \cdot \varepsilon_{ion}$ of other nuclides with the same mass are not negligible, isobaric background is still a problem. This kind of background can be suppressed by resonant laser ionization which is done e.g. at CERN at RILIS (Resonance Ionization Laser Ion Source) [3]. An electron of RI can be excited to a Rydberg state by a laser having exactly the necessary energy and be removed from the atom by a second one. Isobars will not interact with the first laser and the energy of the second laser is not sufficient to put electrons into the continuum. Therefore, only atoms from the desired elements will be ionised. For this separation process a further efficiency $\varepsilon_{sep}$ has to be considered.

- Finally, RI gets accelerated by a second accelerator with the efficiency $\varepsilon_{acc}$.

A schematic overview of these steps and the related efficiencies is shown in fig. 2, which can be seen as a more detailed centre part of fig. 1b).

As many experiments request a high intensity of the secondary beam, we will have a closer look on the relevant parameters. An obvious parameter is the intensity of the primary beam, which is not only limited by the capability of the primary accelerator system but also by the stability of the targets. A well-focussed beam on a thin target rod would be preferable in terms of effusion time out of the target, however, such a target can be destroyed by shock waves when bombarded with a high intensity beam. Therefore, ISOLDE target rods have a diameter of typically 2 cm and a length of about 20 cm. In addition, there are high requirements on radiation protection, not only during operation but also during shutdown due to activation of infrastructure material. For several maintenance steps remote handling is necessary.

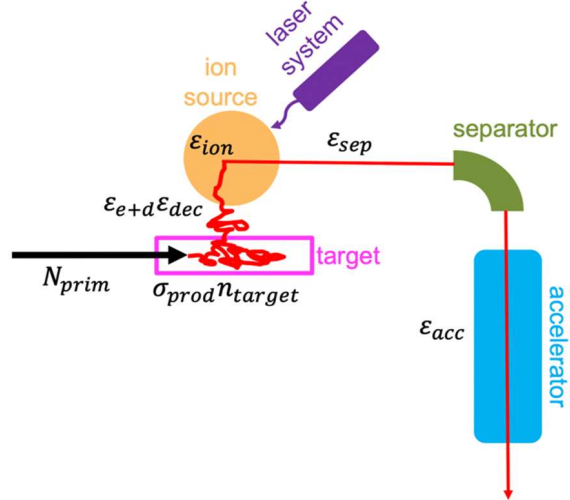

**Fig 2:** Efficiencies for the ISOL process

Another critical parameter is the production cross section $\sigma_{prod}$. A choice of a suitable target element has a large influence on it. Therefore, a large variety of target materials is in use, depending on the desired radioisotope. For radionuclides with a high $\sigma_{prod}$ with neutrons the use of a neutron converter target in direct vicinity of the production target allows an increase of the production yield. In this case, the radionuclide is produced by neutrons originating from spallation reactions in the converter.

In order to increase the efficiency of the effusion, the targets are heated close to melting temperature. At ISOLDE uses typically powders pressed in a tantalum container, at TRIUMF foil stacks are used as target because of their high surface area and short effusion lengths.

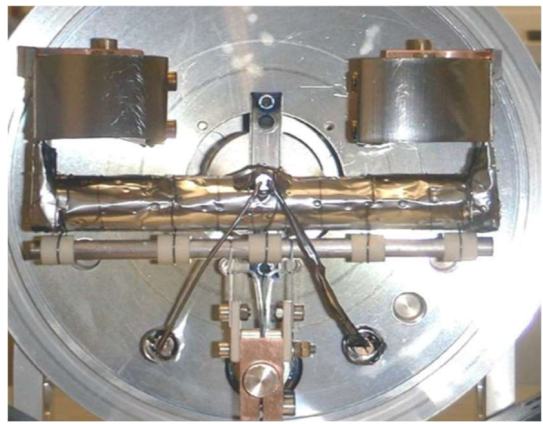

**Fig 3:** ISOL targets from ISOLDE (left) version with neutron converter target made of tantalum below a uranium carbide target.

## 3   The In-Fight Method

A complementary approach to the ISOL method is the so-called In-Flight method which will explained on the example of the fragment separator FRS [4,5] at GSI, Darmstadt, and its upgrade Super-FRS [6], which is presently under construction.

Despite being a very powerful tool for nuclear physics with radioactive ion beam, the ISOL method has two weak points: i) for several chemical element the diffusion (and effusion) efficiency and therefore, their yield is very low, and ii) due to the diffusion time in the order of seconds, the method is not feasible for radionuclides with significantly lower half lives. These shortcuts can be overcome by the In-Flight method, where effects due to chemistry and atomic physics can be neglected and which is considerably faster.

The principle has been already explained in the introduction and is depicted in fig. 1c: a target is bombarded with a heavy ion beam. That part of the projectile nucleus which is overlapping with a target nucleus is sheared off, the remaining part is left relatively unaffected and travels along with the speed of the original projectile. However, this nucleus is in and exited stage and will get rid of the excitation energy mainly by evaporating neutrons immediately. Therefore, the majority of radionuclides produced by fragmentation are neutron deficient. On the other hand, a fissionable projectile like 238U travelling through the electric field of a nearby target nucleus can be fissioned. The fission products will have approximately the same velocity than the original projectile and a similar A/Z, too. As heavy elements are very neutron rich, this holds for the fission fragments as well.

Fig. 4 shows the intensities for a large number of xenon isotopes achievable at the FRS at GSI, already in the last century.

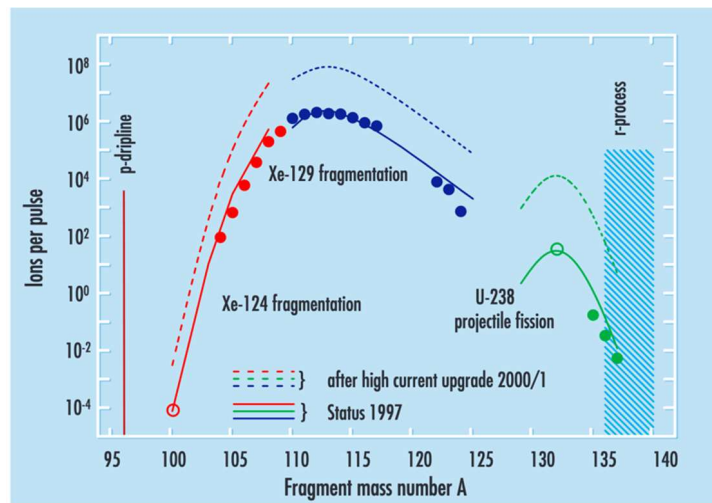

**Fig 4:** Intensities for different xenon isotopes at GSI, achieved with $^{124}$Xe, $^{129}$Xe and $^{238}$U beam, respectively [7].

From a variety of different fragments, the desired radionuclide RI is separated by a special ion optical system like e.g. the FRS (fragment separator) at GSI (see fig. 5). It becomes immediately clear, that there is no long-lasting process step involved. To get an order of magnitude: if we assume 0.3 c for the speed of the fragments and 100 m for the length of the separator, the time required for RI from production target to the experiment is 1 µs. Therefore, beams of radionuclides with half-lives in the µs-range are possible.

After the target, there is a large variety of isotopes with in different charge states, however, with (nearly) the same velocity. In the first half, the fragments are magnetically separated according to their mass to charge ratio A/q. To separate the isotope of interest from this remaining mixture with similar A/q, the particles pass through a thick aluminium degrader. The specific energy loss dE/dx in the degrader depends on the particles nuclear charge Z. Therefore, a further beam purification is done in the separator's seconds stage. In addition, a time-of-flight measurement allows further background suppression. With the GSI's SIS18 synchrotron primary particles are accelerated to energies in the order of 1 GeV/$u$. There are two main advantages of these high energies. Firstly, despite the high energy loss in the thick degrader the energy loss straggling is relatively low. Secondly, lighter fragments a usually completely stripped, thus $A/q$ ambiguities are less problematic.

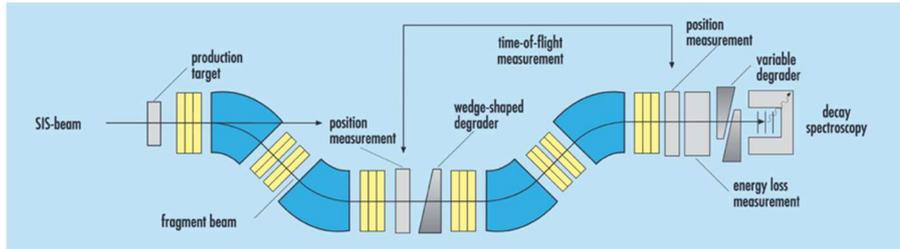

**Fig. 5:** Setup of the FSR at GSI, Darmstadt [7].

There is still a huge interest in understanding stellar nucleosynthesis, in particular the properties of nuclei in the r-process path. In fig. 4 this region is marked for xenon. The relevant nuclei are extremely neutron rich and even for fission fragments the yields are very low. The meet the experimentalist's demands, there are two approaches: The obvious one is to increase the primary beam intensity, a further approach is to increase the acceptance of the separator. Both approaches are followed in the design of the Super-FRS, which is presently under construction at GSI. With FRIB (facility for radioactive ion beams) [8] a comparable facility is under construction at the Michigan State University, East Lansing, USA.

Both approaches a technically (and budgetary) very challenging. For FRS the primary beam for fission is 238U73+, accelerated by the 18 Tm synchrotron SIS18. The intensity is limited by space charge effects. In the future FAIR facility (Facility for Antiproton and Ion Research) SIS18 will be succeeded by the 100 Tm SIS100 synchrotron with a circumference of 1.1 km. With this machine an even higher energy can be reached using a charge states of only around 20+, which will allow an intensity increase by more than two orders of magnitude as shown in fig. 6.

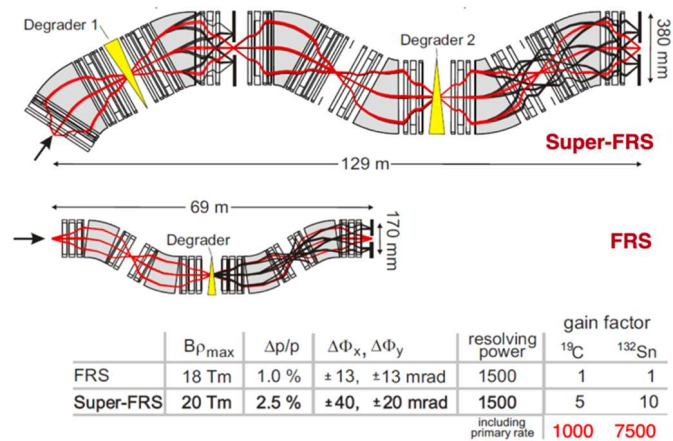

**Fig. 6:** Comparison of FSR and Super-FRS at GSI, Darmstadt [9]

Different fields of physics require precise measurements of the masses of radioactive nuclei, e.g. in astrophysics their knowledge is indispensable for the understanding of stellar nucleosynthesis. Data of extremely high quality can be obtained by storage ring experiments, which are performed e.g. at GSI's Experimental Storage Ring ESR. The principle is illustrated in fig. 7: "The two kinds of mass spectrometry applied at the ESR by measuring the revolution frequencies of stored exotic ions. Left hand side: Schottky mass spectrometry. Here the ions are electron cooled, therefore, their velocity spread

Δv gets negligibly small. Their revolution frequencies are measured by pick up plates mounted in the ring aperture. This technique has been successfully applied at longer-lived exotic nuclei. Right hand side: Isochronous mass spectrometry. Uncooled ions circulate at their transition energy γt. Their revolution times are measured by a time-of-flight technique. This method is in particular suited for short-lived nuclei with half-lives in the millisecond or even microsecond range." [5].

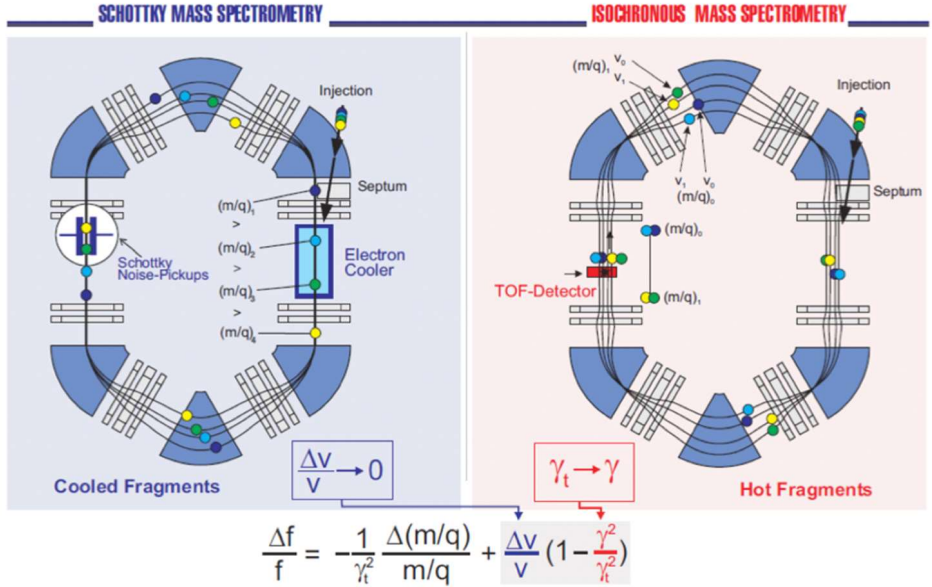

**Fig. 7:** Mass measurements at storage rings [5].

## 4  Secondary beams of exotic particles

Up to now only secondary particles have been discussed, which already "lived" in a larger nucleus from which nucleons have been removed. This chapter is about secondary beams of particles which are created as particle-antiparticle pairs in the target, e.g. antiprotons, pions or kaons. Due to the short lifetime (pions, kaons) or due to the annihilation when in contact with matter (antiprotons) the ISOL method is not feasible. These setups are very similar to those of the in-flight method. In order to explain the principles, I will focus on antiprotons in this chapter.

There has been a large interest on antiproton beams, because of the possibility to operate a single synchrotron like the CERN's SPS or Fermilab's Tevatron as a collider. This can be achieved by operating the machine witch antiprotons and protons at the same time: when protons are rotating in clockwise direction, antiprotons are rotating counter-clockwise and it is possible to collide the beams in designated areas.

Antiprotons (or p̄, pbar) are generated in inelastic collisions of high energy protons with nucleons of a target nucleus at rest by the process pA→p̄X, where X represents all the other particles in any final quantum state allowed in the collision, i.e. the scattered projectile proton, the proton from the p̄-p pair, other hadron or lepton pairs ("shower"), and the residue of the target nucleus with the initial mass number A.

Considering the kinetic energy of the center of mass (c.m.) of the interacting primary nucleons the process requires a minimum kinetic proton energy in the laboratory system above the p̄ threshold of 6 $m_p c^2$ = 5.6 GeV, with $m_p$ as the rest mass of the (anti)proton. The 6.2 GeV (or BeV) Bevatron was built in Berkeley with the main purpose to proof the existence of the antiproton (and antineutron). The successful detection of the antiproton already in 1955 was awarded with the Nobel Prize in 1959. As the production cross section increases with increasing primary beam intensity, higher primary energies are preferable.

The angular distribution of the secondary p̄-p pairs is isotropic in the c.m. system of the two interacting nucleons. Hence, the pbar velocity distribution has its maximum near the c.m. velocity. This isotropy in the c.m. system leads to a relatively wide angular distribution of the antiprotons in the laboratory system as shown in fig. 8.

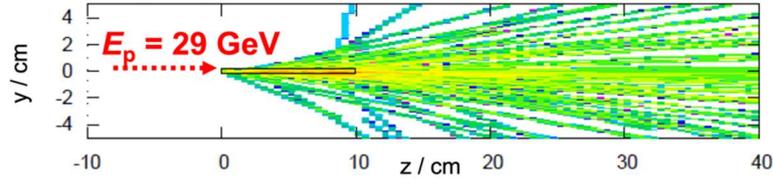

**Fig. 8:** FLUKA simulation [11] of the antiproton distribution after a nickel target bombarded with 29 GeV protons. Note that the scales are not distorted.

For this reason, a strong focusing device directly after the production target is needed. Quadrupole doublets are not sufficient for this purpose. A device which focusses the beam in all directions simultaneously is needed. A magnetic horn [12] or lithium lens [13] can fulfill this purpose. By means of a pulsed current in the order of $10^5 – 10^6$ A a circular magnetic field is produced which is capable of focussing antiprotons with angles of up to 100 mrad with respect to the beam axis.

After the first focussing device the beam quality of the antiproton beam is still very poor. In order to avoid very high losses after the production target, the succeeding beamline needs to have a very high acceptance. E.g. at the future FAIR facility antiprotons with a momentum spread $\Delta p/p$ of ±3 % and an emittance of 240 $\pi$ mm mrad will be accepted. As a consequence, beam tube diameters have to be as large as 400 mm and magnets need to have a very large aperture. Despite this high acceptance, only a few percent of the antiprotons (about $2\times 10^{-5}$ of $10^{-3}$ per primary) can be captured.

The luminosity of the antiproton beam is still very poor at this stage. In order to improve the luminosity significantly, the antiprotons beam is injected in a storage ring, where bunch rotation, adiabatic debunching and stochastic cooling takes place. By these means both emittance and momentum spread can be reduced by one to two orders of magnitude within a few seconds. As the beam size is substantially decreased by these measures, the accumulation of cooled pulses (typically $10^3$) in a succeeding second storage ring becomes possible which results in antiproton beams with reasonable beam qualities and intensities of up to $10^{12}$ particles. More detailed descriptions of these processes would be beyond the scope of this paper, but can be found e.g. in [14]. In table 1 the main parameters of the antiproton sources for the former collider operations at CERN and Fermilab and for FAIR are given.

| Laboratory/Project | CERN | Fermilab | FAIR |
|---|---|---|---|
| Kinetic energy of protons | 25 GeV | 120 GeV | 29 GeV |
| Number of protons per cycle | ~1.5×10¹³ | ~8×10¹² | ~2×10¹³ |
| Cycle time | 4.8 s | 2.2 s | 10 s |
| Kinetic energy of selected p̄ | 2.7 GeV | 8 GeV | 3 GeV |
| Transverse emittance of p̄ beam | 210 mm mrad | 35 mm mrad | 240 mm mrad |
| Momentum spread of p̄ beam | ± 3% | ± 2.25% | ± 3% |
| Overall p̄ yield (per primary p) | ~5×10⁻⁶ | ~2×10⁻⁵ | 5×10⁻⁶-1×10⁻⁵ |

**Tab. 1:** Parameters of the antiproton facilities at CERN [15], Fermilab [16] and the planned FAIR [17] at GSI.

The parameters for CERN and FAIR are very similar, whereas at Fermilab the higher projectile energy results in a significantly higher production cross section, a higher average energy of the antiprotons (requiring a larger ring for collecting the antiprotons) and a lower emittance of the pbar beam.

For all these antiproton sources radiation protection is an additional challenge, mainly because of the high number of secondary neutrons produced in the targets. E.g. at FAIR the concrete walls around the target area will be up to 8 m wide in order to allow access in the neighbouring tunnels during antiproton production.

The principles for the production of pion beams are the same; actually, in all antiproton sources even higher numbers of pions and also kaons are produced. However, the steps performed in the storage rings cannot be applied due to the short lifetime of these particles.

## 5    Tertiary beams

As described in the chapter above, primary particles with an energy above the production threshold will produce secondary pions. For pions in particular, there is the possibility of a production without producing their antiparticles, i.e. the energy threshold is lower. Antiprotons consist of three antiquarks; therefore, it is unavoidable to create twice the number (anti)quarks, i.e. six quarks in total. A positive pion consists of a quark and an antiquark, therefore, a single pion production via $p + p \rightarrow p + n + \pi+$ is possible, too. Only two additional quarks need to be created.

Charged pions have a half-life of only 26 ns and decay to muons and (anti)neutrinos:

$$\pi+ \rightarrow \mu+ + \nu\mu \text{ and } \pi- \rightarrow \mu- + \bar{\nu}\mu.$$

This half-life corresponds to a beam range in the order of $10^2$ m, depending on relativistic corrections, i.e. a beam of secondary pions will finally be converted in a beam of "tertiary" muons and neutrinos.

A large number of experiments with muons is performed e.g. at the Swiss Muon Source SµS at the Paul Scherrer Institute PSI. The secondaries are produced by bombarding a rotating graphite wheel with 590 MeV protons from a high intensity cyclotron delivering 1.4 MW of beam power.

There is also a long history in neutrino beams. For an overview see [18]. The most elaborated applications of accelerator neutrinos are the so-called long baseline experiments like CNGS (CERN Neutrinos to Gran Sasso), Fermilab's NuMI (Neutrinos at the Main Injector) or T2K (Tokai to Kamioka) in Japan.

The principle will be explained on the example of CERN's CNGS: a target consisting of a row of graphite rods was bombarded with 400 GeV protons from the SPS. The secondary π+ beam was parallelised by a set of two large magnetic horns. In a succeeding decay channel of 1 km length a beam of tertiary muon neutrinos had developed which was directed through the Earth to the OPERA detector located in 730 km distance near Rome. This detector was sensitive for tau neutrinos but blind for muon neutrinos. Therefore, it was possible to detect oscillations of the neutrino flavour from νµ to. Although the count rate has been in the order of one ντ event per year, due to the virtually missing background a 5 σ confidence level has been reached after the detection of the 5th tau neutrino.